\documentclass[jkps,preprint,showpacs,floatfix]{revtex4}
\usepackage[dvips]{graphicx}
\begin{document}
\title{Spin-dependent Empirical Formula for the Lowest Excitation Energies of the Natural Parity States in Even-even Nuclei}
\author{Guangho \surname{Jin}}
\author{Dongwoo \surname{Cha}}
\author{Jin-Hee \surname{Yoon}}
\email{jinyoon@inha.ac.kr}
\thanks{Fax: +82-32-866-2452}
\affiliation{Department of Physics, Inha University, Incheon
402-751, Korea}
\date{September 26, 2008}

\begin{abstract}
We present an empirical expression that holds for the lowest excitation energy of the natural parity states in even-even nuclei throughout the entire periodic table. This formula contains spin-dependent factors so that it is applied to different multipole states with the same model parameters, in contrast to the recently proposed empirical expression, for which the model parameters had to be fitted for each multipole separately.
\end{abstract}


\maketitle

Our knowledge of nuclear physics has frequently been increased by phenomenological or empirical studies where some particular nuclear properties have been examined in terms of the simple nuclear variables that govern the abundant experimental data over a wide span of the chart of nuclides. One recent such example is the well-known ``$N_pN_n$ scheme" \cite{Casten}, which denotes the phenomenon of a simple pattern that emerges whenever nuclear data concerning the lowest collective states is plotted against the product $N_pN_n$ between the valence proton number $N_p$ and the valence neutron number $N_n$. The $N_pN_n$ scheme has been very successful for more than two decades in correlating a large amount of data on the collective degrees of freedom in nuclei. Many authors believe that the $N_pN_n$ scheme is manifested because the valence proton-neutron interaction is a dominant controlling factor in the development of collectivity in nuclei \cite{Heyde,Federmann,Dobaczewski}.

Recently, another empirical study, also adopting the valence nucleon numbers, expressed the excitation energy of the lowest collective states in even-even nuclei \cite{Kim,Jin}. In Ref.\,5, the excitation energy of the lowest natural parity even multipole states is given explicitly by the valence nucleon numbers, in contrast to the $N_pN_n$ scheme which says merely that relevant observables can be parametrized in terms of the product $N_pN_n$. More specifically, the empirical formula employed for the excitation energy $E_x$ is given by
\begin{equation} \label{E}
E_x = \alpha A^{-\gamma} + \beta_p  e^{- \lambda_p N_p} + \beta_n
e^{- \lambda_n N_n},
\end{equation}
where $A$ is the mass number of the nucleus and the parameters $\alpha$, $\gamma$, $\beta_p$, $\beta_n$, $\lambda_p$, and $\lambda_n$ are fitted from the data for each multipole. It was shown that this simple formula could describe essential trends of the excitation energy of the lowest natural parity even multipole states in even-even nuclei throughout the entire periodic table \cite{Ha,Kim}. In addition, it was also shown that this empirical formula did, indeed, comply with the $N_pN_n$ scheme even though the formula itself did not explicitly depend on the product $N_pN_n$ \cite{Yoon}. Furthermore, the formula was tested again for the lowest excitation energy of natural parity odd multipole states. The outcome turned out to be reasonably good \cite{Jin}.

However, this formula can be critiqued in that the number of free parameters in the above empirical study is too large, coming to as many as 30, because there are six free parameters, $\alpha$, $\gamma$, $\beta_p$, $\beta_n$, $\lambda_p$, and $\lambda_n$, for each of the five multipoles. This critique is certainly not unreasonable, remembering physicists' old saying that with enough free parameters they can even fit an elephant. In this work, therefore, we want to examine the parameter values obtained from previous empirical studies in order to see whether we can derive a spin-dependent empirical formula employing fewer free parameters in describing the excitation energy of the lowest natural parity states in even-even nuclei.

\begin{figure}[t]
\centering
\includegraphics[width=14.0cm,angle=0]{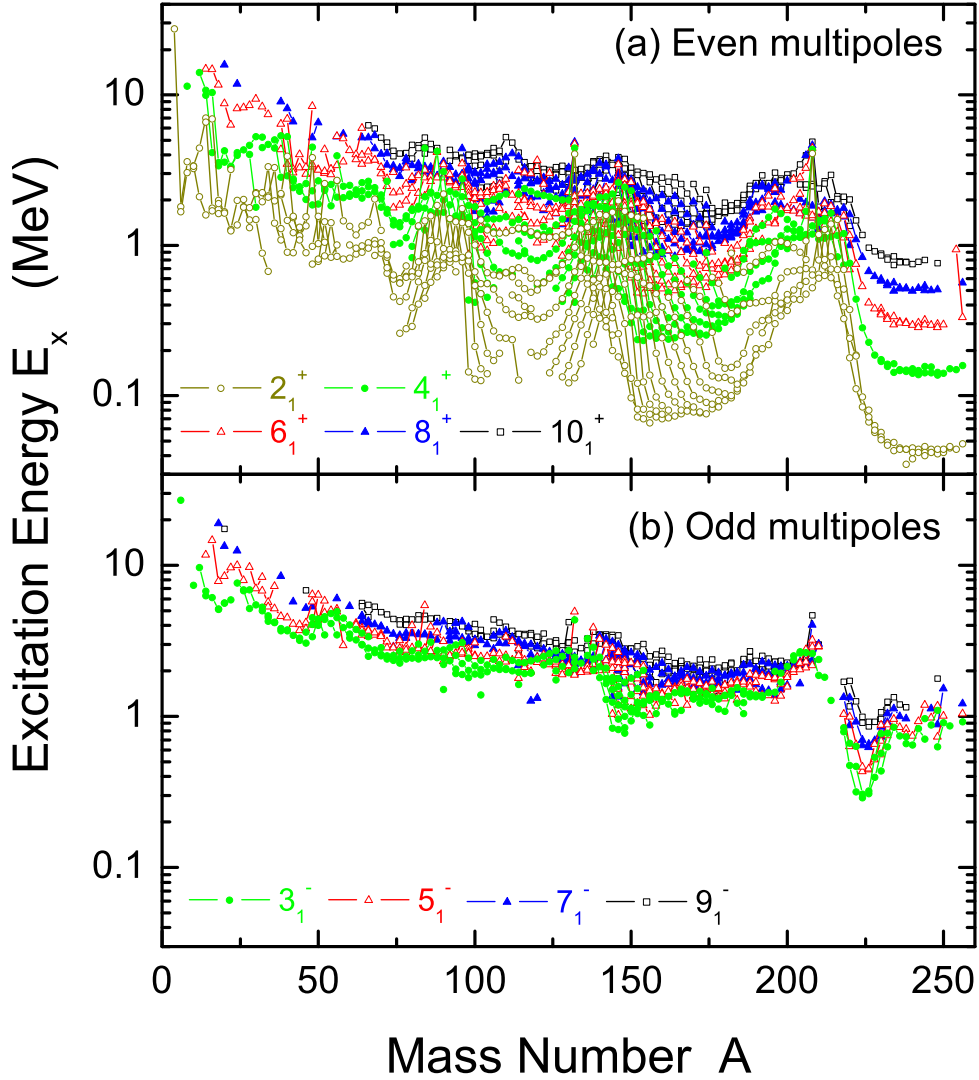}
\caption{Measured excitation energy of the lowest natural parity (a) even multipole and (b) odd multipole states in even-even nuclei \cite{Raman,Firestone,Kibedi}.}
\label{fig-2}
\end{figure}

In order to review the overall shape of the data on the lowest excitation energies in even-even nuclei, we show those for the natural parity even multipole states, including $2_1^+$ (empty circles), $4_1^+$ (solid circles), $6_1^+$ (empty triangles), $8_1^+$ (solid triangles), and $10_1^+$ (empty squares), in Fig.\,\ref{fig-2}(a) while we show those for the natural parity odd multipole states, including $3_1^-$ (solid circles), $5_1^-$ (empty triangles), $7_1^-$ (solid triangles), and $9_1^-$ (empty squares), in Fig.\,\ref{fig-2}(b). The lowest dipole excitation energies $E_x(1_1^-)$ are excluded from Fig.\,\ref{fig-2}(b) because they do not follow the common pattern of the other odd multipole cases shown in Fig.\,\ref{fig-2}(b) \cite{Jin}. In Fig.\,\ref{fig-2}, the plotted points are connected by solid lines along isotopic chains. It is interesting to note by comparing Fig.\,\ref{fig-2} parts (a) and (b) that the gross behaviors, exposed by the data on the lowest excitation energy of the even multipole states and the odd multipole states, are quite different.  Even though the tendency for the excitation energies to  become larger as the multipole of the state increases is still true for both the odd multipole states and the even multipole states, the overall shape of the odd multipole excitation energies is quite different from that of the even multipole excitation energies. In other words, the lowest excitation energies of the odd multipole states lie significantly closer together than those of the even multipole states. Therefore, we seek the spin-dependent parametrization of the empirical formula for the even and for the odd multipole states separately.

\begin{table}[t]
\begin{center}
\caption{Values adopted for the six parameters in Eq.\,(\ref{E}) for the excitation energy $E_x$ of the lowest natural parity states in even-even nuclei. We quoted these parameter values from Ref.\,5 for the even multipole states and from Ref.\,6 for the odd multipole states. The last two columns are the $\chi^2$ value and the total number $N_0$ of the data points, respectively, for the corresponding multipole state.}
\begin{tabular}{crccccc}
\hline\hline
$J_1^\pi$~~~&~~~$\alpha$~~~~&~~~$\gamma$~~~&~~~$\beta_p$($\beta_n$)~~~
&~~~$\lambda_p$( $\lambda_n$)~~~&~~~$\chi^2$&$N_0$\cr
&(MeV)&&(MeV)&&\cr
\hline
$2_1^+$&68.37&1.34&0.83(1.17)&0.42(0.28)&0.126&557\cr
$4_1^+$&268.04&1.38&1.21(1.68)&0.33(0.23)&0.071&430\cr
$6_1^+$&598.17&1.38&1.40(1.64)&0.31(0.18)&0.069&375\cr
$8_1^+$&1438.59&1.45&1.34(1.50)&0.26(0.15)&0.053&309\cr
$10_1^+$&2316.85&1.47&1.36(1.65)&0.21(0.14)&0.034&265\cr
All&&&&&{\bf 0.078}&1936\cr
\hline
$3_1^-$&76.50&0.83&1.07(0.90)&0.40(0.47)&0.073&289\cr
$5_1^-$&144.14&0.92&0.84(1.09)&0.32(0.45)&0.046&297\cr
$7_1^-$&282.54&1.01&0.66(1.08)&0.37(0.56)&0.036&241\cr
$9_1^-$&441.51&1.06&0.77(1.33)&0.32(0.37)&0.022&204\cr
All&&&&&{\bf 0.046}&1031\cr
\hline\hline
\end{tabular}
\label{tab-1}
\end{center}
\end{table}

The parameters $\alpha$, $\gamma$, $\beta_p$, $\beta_n$, $\lambda_p$, and $\lambda_n$ in Eq.\,(\ref{E}) were determined by using a least $\chi^2$ fitting procedure. We defined the $\chi^2$ value through the logarithmic error $R_E (i)$ for the $i$th data point of the calculated excitation energy $E_x^{\rm cal} (i)$ with respect to the measured one $E_x^{\rm exp} (i)$, which was defined by
\begin{equation} \label{error}
R_E(i) =  \log \left[ E_x^{\rm cal}(i) \right] - \log \left[ E_x^{\rm exp}(i) \right].
\end{equation}
Then, the dimensionless $\chi^2$ value was given by
\begin{equation} \label{chi}
\chi^2 = { 1 \over {N_0}} \sum_{i=1}^{N_0} \Big| R_E(i) \Big|^2 ,
\end{equation}
where $N_0$ is the number of total data points considered \cite{Kim}.

\begin{figure}[t]
\centering
\includegraphics[width=11.0cm,angle=0]{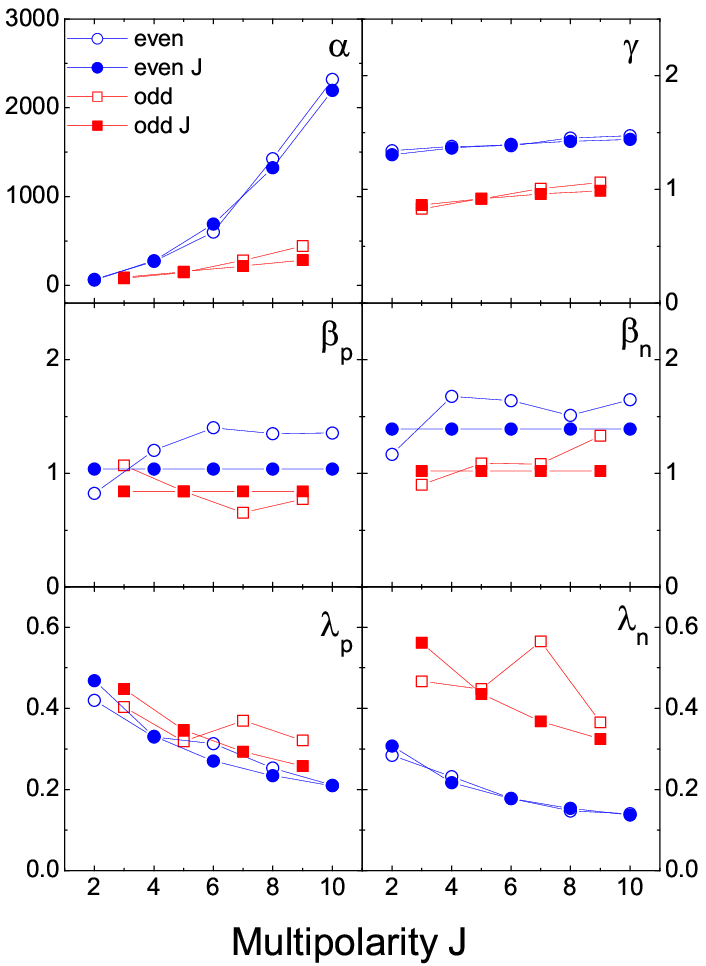}
\caption{Six parameters that appear in Eq.\,(\ref{E}). Open symbols denote the parameter values obtained from previous studies; open circles are for the even multipole states \cite{Kim} while open squares are for the odd multipole states \cite{Jin}. Filled symbols reflect the results obtained by using the spin-dependent empirical formula; filled circles are for the even multipole states while filled squares are for the odd multipole states.}
\label{fig-3}
\end{figure}

The values of the parameters $\alpha$, $\gamma$, $\beta_p$, $\beta_n$, $\lambda_p$, and $\lambda_n$ in Eq.\,(\ref{E}) obtained from previous studies are listed in Table\,\ref{tab-1}, together with the $\chi^2$ values and the total number of data points, $N_0$, for each multipole \cite{Kim,Jin}. The values are also shown in Fig.\,\ref{fig-3} with open circles (even multipoles) and open squares (odd multipoles). The parameter values listed in Table\,\ref{tab-1} were fitted for each multipole separately. The $\chi^2$ values shown on the sixth and the eleventh rows by the heading ``All" represent the overall $\chi^2$ value for all of the even multipole states or for all of the odd multipole states and are obtained by using
\begin{equation} \label{ChiS}
\chi^2({\rm sum}) = \left[ {1 \over {\sum_i N_0 (i)}} \right] \left[ \sum_i N_0(i) \chi^2 (i) \right]
\end{equation}
where $i$ runs separately for all of the even multipoles or for all of the odd multipoles.

\begin{table}[t]
\begin{center}
\caption{Values adopted for the eight parameters in Eq.\,(\ref{sE}) for the excitation energy of the first natural parity even and odd multipole states. The last two columns are the $\chi^2$ value and the total number $N_0$ of the data points, respectively.}
\begin{tabular}{cccccccc}
\hline\hline
$J$~~~&~~~$\alpha_0$~~~&~~~$a$~~~&~~~$\gamma_0$~~~&~~~$c$~~~
&~~~$\beta_p$/$\beta_n$~~~&~~~$\lambda_p^0$/$\lambda_n^0$~~~
&$~~~\chi^2$\cr
&(MeV)&&&&(MeV)&&\cr
\hline
Even&11.94&2.26&1.25&0.06&1.04/1.39&0.66/0.43&{\bf 0.082}\cr
\hline
Odd&28.15&1.05&0.75&0.12&0.84/1.02&0.77/0.97&{\bf 0.048}\cr
\hline\hline
\end{tabular}
\label{tab-2}
\end{center}
\end{table}

By observing the open symbols shown in Fig.\,\ref{fig-3}, we can conjecture how the six parameters $\alpha$, $\gamma$, $\beta_p$, $\beta_n$, $\lambda_p$, and $\lambda_n$ depend on the multipolarity $J$ of the state. Since the first two parameters $\alpha$ and $\gamma$ have the most critical influence on the overall shape of the lowest excitation energies, we want to express them as
\begin{equation} \label{ac}
\alpha=\alpha_0 J^a ~~~{\rm and}~~~ \gamma = \gamma_0 J^c ,
\end{equation}
where $a$ and $c$ are additional parameters introduced to give the proper $J$ dependence. In the case of the two parameters $\beta_p$ and $\beta_n$, we take them as constants with respect to different multipoles $J$ because their previous values fluctuate within a small range without showing any obvious tendency. We take the last two parameters $\lambda_p$ and $\lambda_n$, which depend on $J$, as
\begin{equation} \label{lambda}
\lambda_p = {\lambda_p^0  \over \sqrt{J}} ~~~{\rm and} ~~~ \lambda_n= {\lambda_n^0 \over \sqrt{J}},
\end{equation}
where $\lambda_p^0$ and $\lambda_n^0$ are new $J$-independent parameters that will be fitted in place of $\lambda_p$ and $\lambda_n$, respectively. We bestow such particular dependency on $J$ in Eq.\,(\ref{lambda}) in order to reflect the fact that the previous values of $\lambda_p$ and $\lambda_n$ decreased rather slowly as $J$ became larger, as can be seen in the bottom two panels of Fig.\,\ref{fig-3}.
Finally, then, if Eqs.\,(\ref{ac}) and (\ref{lambda}) are substituted for the parameters in Eq.\,(\ref{E}), the spin-dependent empirical formula now becomes
\begin{equation} \label{sE}
E_x = \alpha_0 J^a A^{-\gamma_0 J^c} + \beta_p  e^{- {\lambda_p^0 N_p \over \sqrt{J}}} + \beta_n e^{- {\lambda_n^0 N_n \over \sqrt{J}}}.
\end{equation}
We use this equation in estimating the logarithmic error $R_E(i)$ defined by Eq.\,(\ref{error}) and in determining the eight parameters $\alpha_0$, $a$, $\gamma_0$, $c$, $\beta_p$, $\beta_n$, $\lambda_p^0$, and $\lambda_n^0$ in Eq.\,(\ref{sE}) by using all of the even or odd multipoles' lowest $E_x$ to minimize the $\chi^2$ value given by Eq.\,(\ref{chi}).

\begin{figure}[t]
\centering
\includegraphics[width=14.0cm,angle=0]{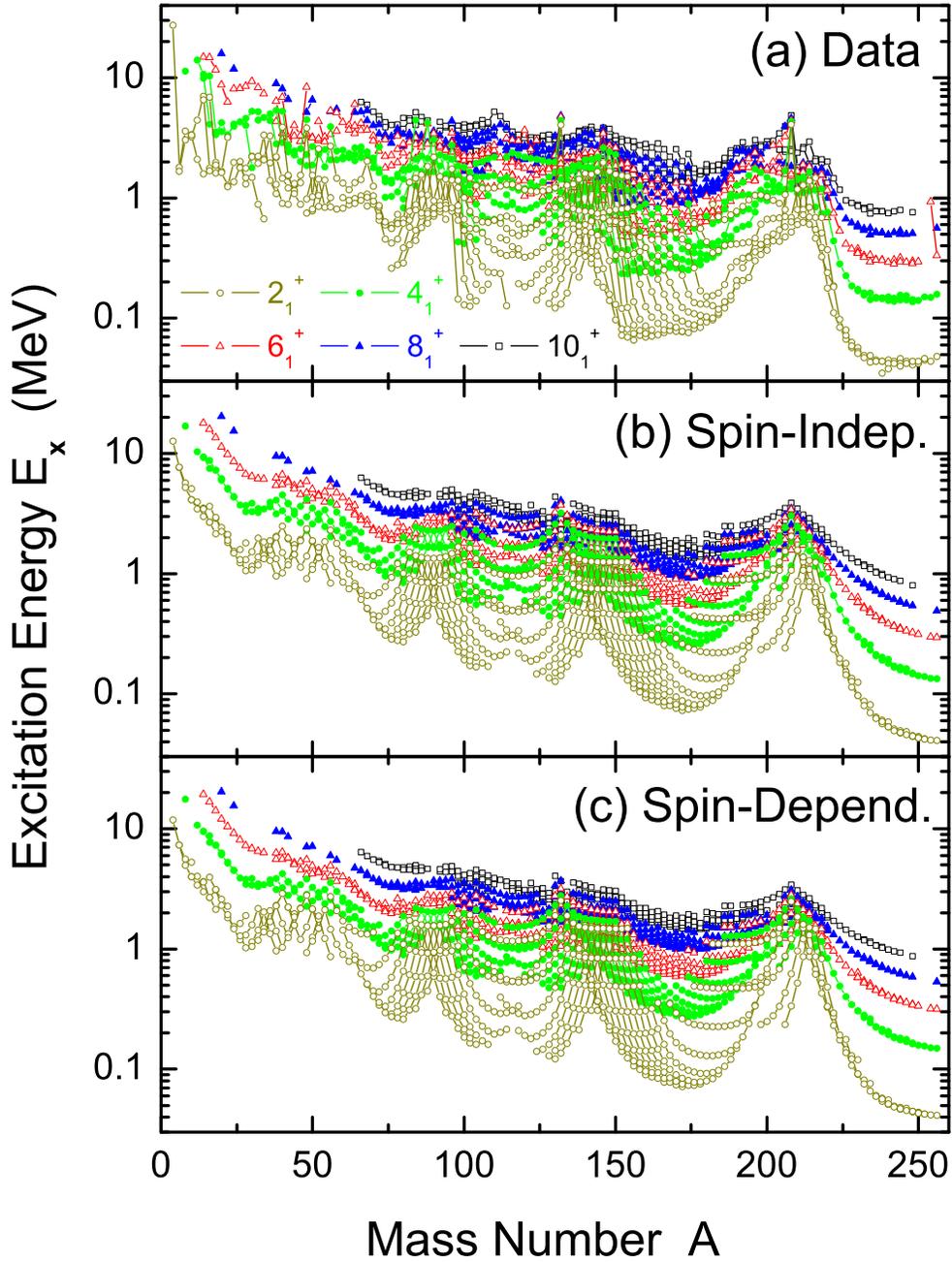}%
\caption{Excitation energies of the lowest natural parity even multipole states up to $10^+$ in even-even nuclei. The plotted points are connected by solid lines along isotopic chains. Part (a) shows the measured excitation energies while parts (b) and (c) show those calculated by using the previous empirical formula (Eq.\,(\ref{E})) and by using the spin-dependent formula (Eq.\,(\ref{sE})).}
\label{fig-4}
\end{figure}

Our results for the eight parameters in Eq.\,(\ref{sE}) are listed in Table\,\ref{tab-2}, together with the $\chi^2$ values and the number of total data points, $N_0$. Also, the values of the six parameters $\alpha$, $\gamma$, $\beta_p$, $\beta_n$, $\lambda_p$, and $\lambda_n$, which are evaluated by using Eqs.\,(\ref{ac}) and (\ref{lambda}), are shown in Fig.\,\ref{fig-3} by filled circles (even multipoles) and filled squares (odd multipoles). By comparing the filled symbols with the corresponding open symbols in Fig.\,\ref{fig-3}, we find that the agreement between the parameters obtained by fitting each multipole separately and those parameters obtained by using the spin-dependent empirical formula is very impressive. Furthermore, by comparing the $\chi^2$ values listed in Table\,\ref{tab-2} with those listed in Table\,\ref{tab-1} for the heading ``All", we find that the increase in the $\chi^2$ value after using the spin-dependent empirical formula is only $\sim 5\%$. This means that the spin-dependent empirical formula, Eq.\,(\ref{sE}), reproduces the previous results of Refs.\,5 and 6 almost exactly. On top of our success in obtaining the spin-dependent empirical formula, we also make an interesting observation: We can infer a distinction in the gross behaviors of the lowest excitation energies between the even and the odd multipole states from the fitted results of the dependence on the multipolarity $J$ of the parameter $\alpha$ in the first $A$-dependent term of Eq.\,(\ref{sE}). According to Fig.\,\ref{fig-3} and Table\,\ref{tab-2}, $\alpha$ depends on $J$ in an almost quadratic manner for even multipoles and an almost linear manner for odd multipoles. These results confirm the well-known fact that most deformed nuclei, with large ground state quadrupole moments, exhibit rotational energy bands for only the even multipoles \cite{Bohr}. In fact, we already pointed out elsewhere that the first term of Eq.\,(\ref{sE}), which reflects the rotational band energies for the mid-shell nuclei, is roughly proportional to $J(J+1)$ when applied to the even multipole states \cite{Jin2}. However, as far as we know, no possible origin for the linear dependency of $\alpha$ on $J$ for odd multipoles has yet been identified.

\begin{figure}[t]
\centering
\includegraphics[width=14.0cm,angle=0]{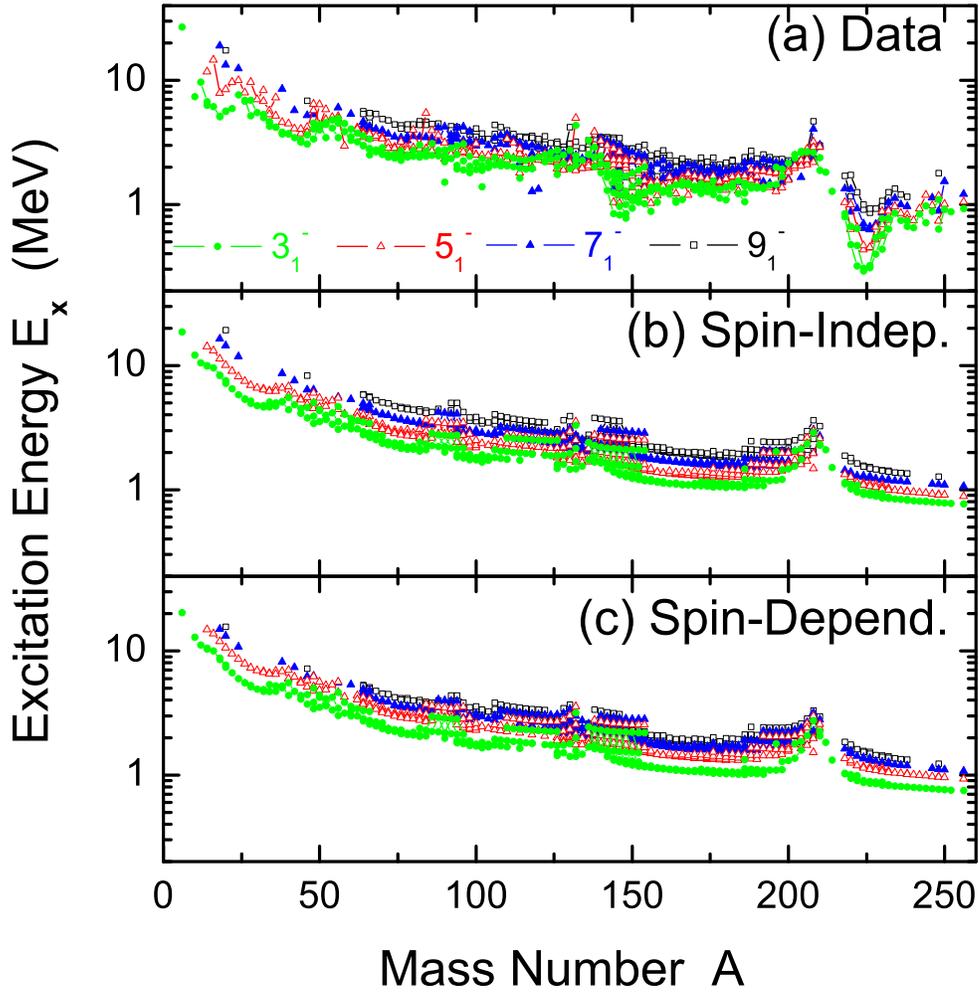}%
\caption{Same as in Fig.\,\ref{fig-4}, but for the lowest excitation energies of the natural parity odd multipole states up to $9^-$ instead of the even multipole states.}
\label{fig-5}
\end{figure}

The performance of the spin-dependent empirical formula can be confirmed more explicitly by inspecting Fig.\,\ref{fig-4} for the even multipole states and Fig.\,\ref{fig-5} for the odd multipole states. In Fig.\,\ref{fig-4}, the excitation energies of the lowest natural parity even multipole states, up to $10^+$ in even-even nuclei, are shown in three panels. The top panel, part (a), shows the measured excitation energies while the middle and the bottom panels, parts (b) and (c), show those calculated by using the previous empirical formula (Eq.\,(\ref{E})) and by using the current spin-dependent one (Eq.\,(\ref{sE})), respectively. The graph shown in part (a) of Fig.\,\ref{fig-4} is exactly the same as that of Fig.\,\ref{fig-2}(a). We find that the graph shown in part (b) of the same figure reproduces the essential trend of the measured excitation energies shown in part (a), as discussed in Ref.\,5. Now, it is really remarkable to find that the graph shown in part (c) of Fig.\,\ref{fig-4} is almost exactly the same as that shown in part (b). Therefore, we conclude that the spin-dependent empirical formula, Eq.\,(\ref{sE}), can be used in place of the previous empirical formula, Eq.\,(\ref{E}), for even multipole cases.

The same argument discussed above with respect to Fig.\,\ref{fig-4} for the even multipole states can also be applied to the graphs shown in Fig.\,\ref{fig-5} for the odd multipole states, including $3^-$, $5^-$, $7^-$, and $9^-$ in even-even nuclei. However, as is evident by comparing the graphs shown in parts (a) and (b), there exist two particular groups of nuclei where the measured lowest excitation energies do not follow the overall trend imposed by the previous empirical formula, Eq.\,(\ref{E}), between the mass numbers, $A=144 \sim 152$ and $A=220 \sim 232$. According to Ref.\,6, such a group of nuclei happens to be all of the isotopes whose neutron number is equal to an even number from 86 to 92 (between $A=144 \sim 152$) or all of the isotopes whose proton number is equal to an even number from 86 to 92 (between $A=220 \sim 232$). If those two groups of nuclei are excluded from consideration, we can see that the graph shown in part (b) of Fig.\,\ref{fig-5} also reproduces the essential trend of the measured excitation energies shown in part (a). Finally, we can observe that the graphs shown in parts (b) and (c) are almost identical.

In summary, we have shown that a new spin-dependent parametrization can be devised for the empirical formula and that the new parametrization holds for the lowest excitation energy of the natural parity states in even-even nuclei throughout the entire periodic table. This is certainly meaningful progress compared to the previous such empirical formula in the sense that the old one should be fitted for each multipole state separately.

\begin{acknowledgments}
This work
was supported by an Inha University research grant.
\end{acknowledgments}

\end{document}